\documentclass[aps,floatfix,notitlepage, tightenlines, nofootinbib]{revtex4-1}
\usepackage{amsmath, amssymb, amsfonts, amsthm, latexsym, epsfig, mathrsfs, xcolor, bbm, slashed}
\usepackage[top=1in, bottom=1in, left=.9in, right=.9in]{geometry}
\DeclareMathAlphabet{\mathpzc}{OT1}{pzc}{m}{it}
\usepackage[cal=boondox, calscaled=.94, bb=ams, frak=mma, frakscaled=.97, scr=rsfs]{mathalfa} 
\usepackage[T1]{fontenc} 
\usepackage{amsfonts}
\usepackage{graphicx}
\usepackage{amsmath}
\usepackage{color}
\usepackage{tasks}
\settasks{
	counter-format=(tsk[r]),
	label-width=4ex
}
\usepackage[colorlinks=true, citecolor=blue,urlcolor=blue]{hyperref}
\usepackage{hyperref}
\usepackage{PeterStyle}
\usepackage{subfigure}
\usepackage{comment}
\usepackage{multirow}
\usepackage{calligra}

\newcommand{\PZ}[1]{\textcolor{red}{\textsf{[PZ: #1]}}}
\newcommand{\SG}[1]{\textcolor{blue}{\textsf{[SG: #1]}}}

\def\Us{\ensuremath{\Upsilon_{s}}}
\def\hUs{\ensuremath{\hat \Upsilon_s}}
\def\s{\ensuremath{\mathcal{s}}}
\def\tUs{\ensuremath{\tilde \Upsilon_{s}}}
\def\tUslm{\ensuremath{{}_s \tilde \Upsilon_{\ell m}}}
\def\tf{\ensuremath{{}_s \tilde g_{\ell m\omega }}}
\def\htf{\ensuremath{{}_s \hat{\tilde {g}}_{\ell m\omega }}}
\def\Sh{\ensuremath{{}_s S_{\ell m \omega}}}
\def\Shz{\ensuremath{{}_s S_{\ell m }}}
\def\Rh{\ensuremath{ {}_s R_{\ell m \omega }}}
\def\Rfhz{\ensuremath{ {}_s R_{\ell m  }}}
\def\Ih{\ensuremath{ {}_s I_{\ell m \omega}}}
\def\Ihz{\ensuremath{ {}_s I_{\ell m}}}
\def\Khz{\ensuremath{ {}_s K_{\ell m}}}
\def\Eh{\ensuremath{ {}_s  \mathcal{E}_{\ell m \omega}}}
\def\hF{\ensuremath{{}_s F_{\ell m}}}
\def\hU{\ensuremath{{}_s \hat{\Upsilon}_{\ell m}}}
\DeclareMathOperator{\thorn}{\text{\rm \th}}
\let\eth\relax
\DeclareMathOperator{\eth}{\text{\rm \dh}}
\newcommand{\Lie}{\pounds} 

\begin{document}

\title{Critical Exponents of Extremal Kerr Perturbations}
\author{Samuel E.~Gralla\footnote{{\tt sgralla@email.arizona.edu}}}
\author{Peter~Zimmerman\footnote{{\tt peterzimmerman@email.arizona.edu}}}
\affiliation{Department of Physics, University of Arizona}
\date{\today}

\begin{abstract}
We show that scalar, electromagnetic, and gravitational perturbations of extremal Kerr black holes are asymptotically self-similar under the near-horizon, late-time scaling symmetry of the background metric.  This accounts for the Aretakis instability (growth of transverse derivatives) as a critical phenomenon associated with the emergent symmetry.  We compute the critical exponent of each mode, which is equivalent to its decay rate. 
It follows from symmetry arguments that, despite the growth of transverse derivatives, all generally covariant scalar quantities decay to zero.
\end{abstract}

\maketitle

\section{Introduction and summary}
Given decades of work indicating the basic stability of four-dimensional, asymptotically flat black holes to massless perturbing fields 
(e.g. \cite{ReggeWheeler1957,Kay:1987ax, Whiting:1988vc,Dafermos:2014cua,Dias:2015wqa}), Aretakis' 2010 discovery of a horizon instability of extremal black holes \cite{Aretakis:2010gd,Aretakis:2012ei} came as something of a surprise.  The instability has a rather unusual  character: the growth is polynomial (rather than exponential), occurring only on the horizon,  and only for sufficiently high-order transverse derivatives of the perturbing field.  Off the horizon, the field and all its derivatives decay to zero \cite{Aretakis:2010gd,aretakis2012decay}. On the principle that there are no accidents in physics, one naturally seeks a deeper explanation: What is the origin of this peculiar behavior?

The answer, as usual, is symmetry.  We will show how the Aretakis instability can be viewed as a \textit{critical phenomenon} associated with the  emergence of a scaling symmetry near the horizon.  The relevant scaling limit \cite{Bardeen:1999px} is well-known for its near-horizon character, but we observe that it also entails late times and is therefore naturally suited to questions of late-time behavior on the horizon.  We show that perturbing fields are asymptotically equal to a sum of terms that are self-similar in the limit.  
The self-similarity accounts for the detailed structure of the instability (transverse derivatives modifying the late-time behavior by one positive power of time), with the numerical values of the exponents completing the description with detailed decay and growth rates.

A second question for the Aretakis instability concerns its physical consequences.  In previous work with A. Zimmerman \cite{Gralla:2016sxp}, we emphasized definite consequences for particular observers or particles, without addressing  observer-independent quantities.
Very recently, independent work of Hadar and Reall \cite{Hadar:2017ven} and Burko and Khanna \cite{Burko:2017eky} argued, from different angles using different techniques, that general covariance prevents such quantities from becoming large.
Burko and Khanna \cite{Burko:2017eky} numerically confirmed the rates \cite{Casals:2016mel,Gralla:2016sxp} for scalar ($\phi \sim v^{-1/2}$) and gravitational ($\psi_4 \sim v^{3/2}$ and $\psi_0 \sim v^{-5/2}$) perturbations of Kerr\footnote{We derived the extremal, spin-$0$ rates in work with M. Casals \cite{Casals:2016mel}, where we also mentioned (without derivation) the rate for $\psi_4$ in the Hartle-Hawking tetrad.  We then derived the complete spin-$s$ non-axisymmetric \textit{near}-extremal rates in work with A. Zimmerman \cite{Gralla:2016sxp}, noting that they agree with the results of our (still unpublished) derivation in the precisely extremal case.  These rates were subsequently confirmed by Refs.~\cite{Richartz:2017qep} and \cite{Burko:2017eky} using different techniques.  In this paper we at last publish the derivation of the rates in the precisely extremal case, together with additional detail about self-similarity as well as (previously unreported) axisymmetric rates.  All of this work is limited to fields arising from compactly supported initial data not extending to the event horizon.} and noted that the polynomial non-derivative invariants of the Riemann tensor are determined from $\psi_0 \psi_4$, which decays like $1/v$.  They gave further examples of scalar invariants that decay, showing how growing and decaying factors always balance in covariant expressions.  Hadar and Reall \cite{Hadar:2017ven} systematized this type of argument in the context of effective field theory, giving an elaborate demonstration of the cancellations in a very general setting.  

In this paper we will use the scaling symmetry to vastly streamline such arguments.  Each self-similar tensor is assigned a scaling weight which obeys simple rules under tensorial operations.  The decay rate of a scalar is the real part of its weight and hence can be determined immediately from the weights of its tensor constituents.  We compute the weights (critical exponents) for modes of scalar, electromagnetic, and gravitational perturbations arising from generic initial data compactly supported away from the horizon.  The exponents all have negative real parts (the largest is $-1/2$), immediately implying the decay of all scalars constructed covariantly from the field and its derivatives.  This confirms recent suggestions that all scalar invariants decay and demonstrates the power of the emergent symmetry as an organizing principle.  These calculations also offer tantalizing hints of the universality that normally accompanies critical phenomena.

In Sec.~\ref{sec:scaling} we discuss 
the mathematics of self-similar tensors and show how it accounts for the Aretakis instability while  streamlining the computation of rates.  In Sec.~\ref{sec:computation} we demonstrate self-similarity for Kerr perturbations and compute the critical exponents by finding the near-horizon, late-time asymptotics of the mode Green functions in closed form.  In Sec.~\ref{sec:outlook} we discuss future directions.  Our conventions are $G=c=1$. We go on to set $M=1$ in our calculations of decay rates.
\section{The Aretakis instability as a critical phenomenon}\label{sec:scaling}
Consider the extremal Kerr metric in ingoing Kerr coordinates $(v,r,\theta,\varphi)$  and introduce
\begin{align}\label{eq:corotating}
 x = \frac{r-M}{M},  \quad \Phi = \varphi - \frac{v}{2M}.
\end{align}
In these ``horizon corotating'' coordinates $(v,x,\theta,\Phi)$, the future event horizon is $x=0$ and its generators have constant $\theta$ and $\Phi$.  Now introduce a scaling parameter $\lambda$ and corresponding scaling coordinates $\bar{x}^\mu$ as
\begin{align}
\label{scaling}
	\bar{v}=\frac{\lambda v}{2M},\quad
	\bar{x}=\frac{x}{\lambda},\quad
	\bar{\theta}=\theta,\quad
	\bar{\Phi}=\Phi.
\end{align}
Changing to these coordinates and letting $\lambda \rightarrow 0$ produces the ``near-horizon extremal Kerr'' (NHEK) metric \cite{Bardeen:1999px},
\begin{align}\label{NHEK}
ds^2 = 2 M^2 \Gamma(\bar\theta) \Big( - \bar x^2 d{\bar v}^2 + 2 d{\bar v} d{\bar x} + d\bar \theta ^2 + \Lambda^2(\bar \theta) \left(d \bar\Phi+ {\bar x} d{\bar v} \right)^2 \Big),
\end{align}
where $\Gamma(\bar\theta)= (1+\cos^2\bar\theta)/2$ and $\Lambda(\bar\theta)= \sin\bar\theta/\Gamma(\bar\theta)$.
Notice that the coordinate transformation \eqref{scaling} is invariant under $\{\lambda \rightarrow c \lambda, \ v\to v/c, \ x \to cx \}$.  Since $\lambda$ disappears in the $\lambda \rightarrow 0$ limit, the resulting metric must enjoy the symmetry $\{\bar{v}\to \bar{v}/c, \bar{x}\to c\bar{x}\}$.  Indeed, this scaling symmetry is manifest in \eqref{NHEK}, and the new Killing vector is the  ``dilation'' $H_0$,
\begin{align}\label{H_0}
H_0 = \bar{v} \pd_{\bar{v}} - \bar{x} \pd_{\bar{x}} = v \pd_{v} - x \pd_x.
\end{align}
The NHEK metric in fact has a second additional Killing field, an analog of global time-translation in Anti-deSitter space.  The second symmetry enhancement is in a sense accidental as it does not follow directly from the limit, though see \cite{Kunduri:2007vf} for the robustness of this feature.  The four Killing fields together form the algebra $\mathrm{sl}(2,\mathbb R)\times \mathrm{u}(1)$, a property sometimes called a ``global conformal symmetry''. 
There are also associated local conformal symmetries \cite{Guica:2008mu}.  In this paper we will consider only the dilation \eqref{H_0}, whose action we refer to as ``scaling''.

\begin{figure}
\includegraphics[scale=.5]{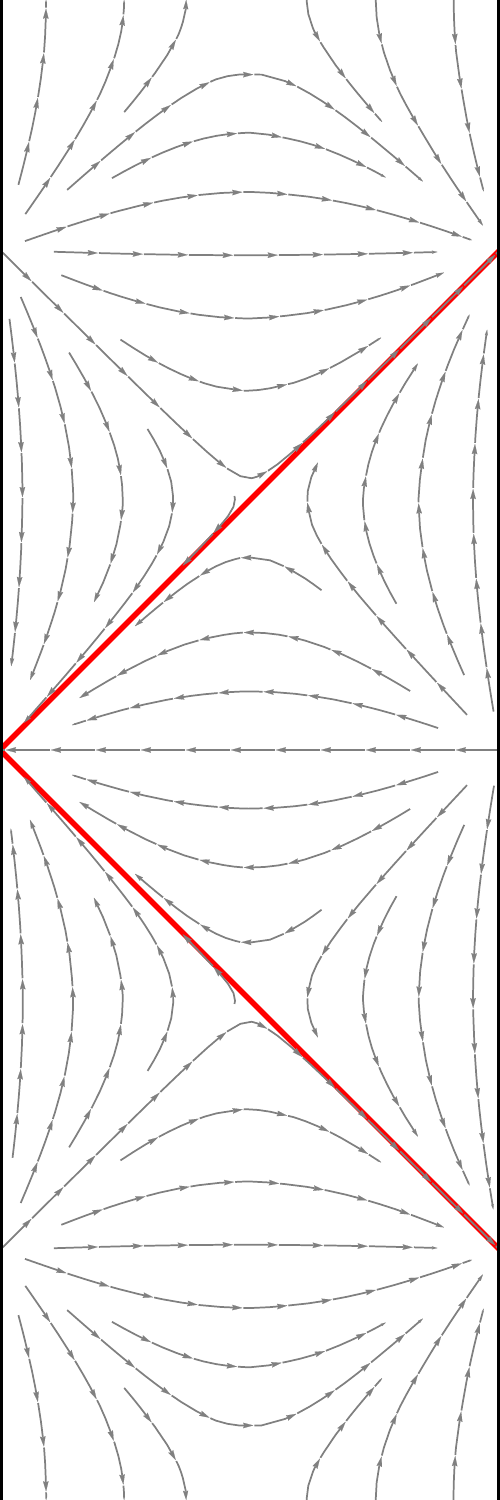}
\caption{The NHEK spacetime, plotted in global coordinates \cite{Bardeen:1999px}. The dilation Killing field $H_0$ is shown in gray, and the Poincar\'e horizon is shown in red.  The spacetime arises in a scaling limit from extremal Kerr, which flows to infinite affine parameter along $H_0$. Correspondingly, tensor fields become self-similar under $H_0$, giving rise to the Aretakis instability.
}\label{fig:NHEK}
\end{figure}

Previous work involving one of us \cite{Gralla:2016jfc} generalized the notion of symmetry enhancement to tensor fields besides the metric.  In particular, Ref.~\cite{Gralla:2016jfc} considered smooth tensor fields on the extremal Kerr exterior (including the horizon) whose barred coordinate components asymptotically\footnote{In this section, when an equation is followed by ``as ...'', we mean that only the leading term is kept [i.e., asymptotic equality in the sense of Poincar\'e, normally denoted with a tilde ($\sim$)].} satisfy
\begin{align}\label{W}
W = \lambda^{-p} \bar{W} \textrm{ \ \ \ \ as $\lambda \to 0$ fixing $\bar{x}^\mu$},
\end{align}
where the (barred-coordinate) components of $\bar{W}$ are independent of $\lambda$. 
This implies that $\bar{W}$ can be considered a 
tensor field in NHEK.  Note that the unbarred coordinates cover the extremal Kerr exterior including the future event horizon $x=0$ (but not including the past event  horizon).  The future horizon is a fixed surface of the scaling limit, where $\lambda \rightarrow 0$ flows along the horizon generators to late times.  The future event horizon $x=0$ limits to the null surface $\bar{x}=0$ in NHEK, which is called the future Poincar\'e horizon.  
The maximally extended NHEK spacetime is shown in Fig.~\ref{fig:NHEK}.

Eq.~\eqref{W} parallels the expansion of fields near critical points in condensed matter physics.  
It follows by similar arguments as given for the metric above that $\bar{W}$ is invariant under $\{\bar{v}\to \bar{v}/c, \ \bar{x}\to c\bar{x}, \
\bar{W} \to c^p\bar{W} \}$.  Infinitesimally, the field satisfies 
\begin{align}\label{emergent}
\Lie_{H_0} \bar{W} = p \bar{W}.
\end{align}
That is, fields of the form \eqref{W} have an emergent scaling self-similarity with weight $p$.  Ref.~\cite{Gralla:2016jfc} studied the consequences of this symmetry for stationary, axisymmetric electromagnetic fields, noting that it accounts for the black hole Meissner effect \cite{Bicak1985}.  The metric has weight $p=0$, corresponding to a true symmetry rather than just self-similarity.

In this paper we study the consequences of the emergent self-similarity \eqref{emergent} for generic (nonstationary, nonaxisymmetric) tensor fields.  We immediately notice that the scaling limit \eqref{scaling} flows not only to the horizon $x \rightarrow 0$, but also to late times $v\rightarrow \infty$.\footnote{If $\bar{v}<0$ the limit instead flows to early advanced times.  This region is not relevant for initial data supported outside the horizon.  The special case $\bar{v}=0$ flows to a fixed point of the future horizon.  Henceforth we assume $\bar{v}>0$.} Thus the scaling is precisely suited to study the Aretakis instability.  We begin with a (possibly complex) scalar $W=\Psi$ for simplicity.  In this case the general solution of Eq.~\eqref{emergent} is
\begin{align}\label{Phiform}
\bar{\Psi} = \bar{v}^p f(\bar{x} \bar{v},\bar{\theta},\bar{\Phi}),
\end{align}
where $f$ is a smooth function of its arguments (and $2\pi$-periodic in the last one).\footnote{The smoothness of $\Psi$ as a scalar field on extreme Kerr requires the smoothness of $\bar{\Psi}$ as a scalar field in NHEK (at least for $\bar{v}>0$).  If we instead wrote $\Psi=\bar{x}^p f(\bar{x}\bar{v},\bar{\theta},\bar{\Phi})$, then  $f$ would not in general be a smooth function of its arguments.} Noting that $\lambda \rightarrow 0$ (at fixed barred coordinate) corresponds to $v \to \infty$ and $x \to 0$, Eq.~\eqref{W} becomes
\begin{align}\label{late}
\Psi = v^p f\left(\frac{xv}{2M},\theta,\Phi \right) \textrm{ \ \ \ \ as $v \to \infty, \ x \to 0$}.
\end{align}
The scaling parameter $\lambda$ has now disappeared from the description, leaving a prediction for the near-horizon, late-time behavior of the original field $\Psi$.  Since $f$ is smooth in its arguments, we see that the decay/growth on the horizon is given by
\begin{align}\label{Aretakis}
(\pd^n_x \Psi)|_{\mathcal{H}} = C_n(\theta,\Phi) v^{p+n}, \qquad \mathrm{as \ \ } v \rightarrow \infty,
\end{align}
where $\mathcal{H}$ means evaluation on the horizon $x=0$.  Here $C_n$ (related to derivatives of $f$ in its first argument) is a constant along each generator labelled by $\theta,\Phi$.  Thus we see that the decay rate is the real part of the weight $p$ (with logarithmic oscillations if $p$ is complex), where  successive transverse derivatives decay slower and ultimately grow unboundedly in time.  That is, the precise Aretakis form is obtained from the natural scaling assumption \eqref{W}, with full details available once the weight $p$ is calculated.  

We may repeat this analysis for a general tensor of arbitrary (finite) rank (App.~\ref{appendix}).  Each component  satisfies Eqs.~\eqref{Phiform}, \eqref{late} and \eqref{Aretakis} except with $p \rightarrow p'=p+N$, where $N=({\rm \# \,of\, upper}\, v\, {\rm indices}-{\rm \#\, of\, upper}\,x\, {\rm indices})+ ({\rm \# \,of\, lower}\, x\, {\rm indices}-{\rm \#\, of\, lower}\,v\, {\rm indices})$.  For example, the component $W^{xvx}{}_{\theta \varphi}$ of a rank-5 tensor $W^{\alpha \beta \gamma}{}_{\mu\nu}$ has $N=-1$ and decays like $v^{p-1}$. 
Thus different components decay/grow at different rates in a manner precisely controlled by $p$.  The weight $p$ provides a coordinate-independent notion of the growth/decay rate.
\subsection{Strong and Weak Self-similarity}\label{sec:rules}
When terms in the late-time expansion of tensor fields with different weights are added together, only the term with larger (real part of) weight survives the asymptotic limit.  If the real parts are the same, then both terms must be kept.  We now formalize this idea by introducing notions of strong and weak self-similarity.  A field with definite weight is strongly self-similar, while a field that is a sum of terms with the same real part of weight is weakly self-similar.  Both strongly and weakly self-similar fields have decay/growth rates controlled by their weight.

We say that a tensor field in Kerr is \emph{strongly self-similar} if it has a definite weight $p$, i.e., if it can be expressed in the form of Eq.~\eqref{W}.
Correspondingly, a tensor field in NHEK is strongly self-similar, with definite weight $p$, if it satisfies Eq.~\eqref{emergent}.  By either definition, it follows that
\begin{tasks}
\task The weight is preserved by scalar multiplication, contraction, and (metric-compatible) derivation.
\task  The tensor product of a tensor of weight $p$ with a tensor of weight $q$ gives a tensor of weight $p+q$.
\task Addition of two tensors with the same weight produces a tensor of the same weight.
\end{tasks}
Using the notion \eqref{emergent} of weight in NHEK, adding fields of different weights breaks the self-similarity property: the resulting field does not have definite weight.  However, referring back to the more fundamental definition \eqref{W}, we see that adding fields of different weights produces a field with weight equal to the input weight with the larger real part.  To deal with the case where the real parts are equal we introduce the notion of weak self-similarity as follows,
\begin{itemize}
\item (\textit{def.}) A tensor $W$ is said to be \emph{weakly self-similar} of weight $P\in \mathbb{R}$ if it is asymptotically equal to a sum of terms of definite weights $p_i$, all with the same real part $P=\textrm{Re}[p_i]$.  That is, $W=\sum_{i} \lambda^{-p_i} \bar{W}_i$ asymptotically as $\lambda \rightarrow 0$ fixing $\bar{x}^\mu$. 
\end{itemize}
In particular, every strongly self-similar tensor of weight $p$ is also weakly self-similar with weight $P=\textrm{Re}[p]$.  The weight of   weakly self-similar tensors satisfies properties (i)-(iii) above as well as
\begin{tasks}[resume=true]
\task The sum of weakly self-similar tensors with weights $P_1$ and $P_2$ is weakly self-similar with weight equal to the larger of $P_1$ and $P_2$.
\end{tasks}
The weight $P$ of a weakly self-similar tensor provides its horizon decay rate in the same manner as strongly self-similar tensors. 
More precisely, a weakly self-similar scalar $\Psi$ satisfies\footnote{An important caveat is that this decay estimate is only guaranteed to hold if the field is a \textit{finite} sum of definite-weight terms.  If the sum is infinite, there is potential for  non-uniformity in the asymptotic approximations to give rise to late-time behavior that differs from that of individual terms.  In particular, our decay estimates have only been established for individual modes, and in principle the behavior of the summed field could be different.}
\begin{align}
\abs{\pd_x^n \Psi}_{\mathcal{H}} \leq C_n v^{P+n}
\end{align}
for some constants $C_n$.  Similarly, each corotating-coordinate component \eqref{eq:corotating} of a weakly self-similar tensor $W$ satisfies the analogous equation with $P \to P'=P+N$, where, as before, $N=({\rm \# \,of\, upper}\, v\, {\rm indices}-{\rm \#\, of\, upper}\,x\, {\rm indices})+ ({\rm \# \,of\, lower}\, x\, {\rm indices}-{\rm \#\, of\, lower}\,v\, {\rm indices})$.  

We have seen that the weight (weak or strong) of a tensor completely characterizes the scaling of the field, and associated scalar observables, in the near-horizon, late-time limit.   Viewing this limit as tuning to a critical point where fields are precisely self-similar, we refer to the weights as critical exponents of extremal Kerr perturbations.  These may be related to critical exponents in the usual sense of taking the temperature to a critical value (in this case absolute zero) by considering simultaneous near-horizon, near-extremal limits.  We give some discussion in Sec.~\ref{sec:outlook} below, but in general leave this to future work.
\subsection{Summary of Results}
In Sec.~\ref{sec:computation} below we consider scalar, electromagnetic, and gravitational perturbations of extremal Kerr that arise from initial data exterior to the event horizon.  That is, we consider a scalar field $\Psi$ satisfying the massless Klein-Gordon equation, a vector field $A_\mu$ satisfying the vacuum Maxwell equation, and a symmetric tensor field $h_{\mu \nu}$ satisfying the vacuum linearized Einstein equation.  Out of the sea of details arises a strikingly simple result: these fields are weakly self-similar (up to gauge) with the universal weight
\begin{align}\label{eq:boss}
P = -1/2.
\end{align}
One may now compute the weight, and hence decay or growth, of any associated scalar of tensor using the algebraic rules for manipulating weights (Sec.~\ref{sec:rules}).  In particular, since the Kerr metric is of weight $0$, it follows immediately that all scalars constructed from the field and metric (including derivatives) decay at least as fast as $v^{-1/2}$.  Computing the decay rate of a scalar involves simply counting the number of times a field appears.  For example, any scalar quadratic in the field (e.g. the field strength invariant $F_{\mu \nu}F^{\mu \nu}$) will decay like $v^{-1}$.  Similarly, the perturbed Kretschmann invariant $R_{\mu \nu \rho \sigma}R^{\mu \nu \rho\sigma}$ will decay like $v^{-1}$ up to gauge.  To give a higher-derivative example, the linear perturbation of $R^{\alpha \beta \gamma \delta} \nabla_\alpha \nabla_\gamma R_{\mu \nu \rho \sigma} \nabla_\beta \nabla_\delta R^{\mu \nu \rho\sigma}$ will decay like $v^{-3/2}$ up to gauge.   One can similarly compute the decay rate for a component of a tensor by first determining its weight and then using the rules given in the last paragraph before Sec.~\ref{sec:rules}.

One can also compute rates for fields built from additional structure besides the metric, provided the weight of that structure is known.  For example, the linear perturbation of the Weyl scalar $\psi_4=C_{\mu \nu \rho \sigma}n^\mu \bar{m}^\nu n^\rho \bar{m}^\sigma$ is built from the (background) metric $g_{\mu \nu}$, its derivative operator $\nabla_\mu$, the metric perturbation $h_{\mu \nu}$, and the Newman-Penrose tetrad $\{\ell^\mu, n^\mu, m^\mu, \bar{m}^\mu\}$.  If the tetrad is weight-$0$ (like the hatted tetrad introduced in Sec.~\ref{sec:reg} below), then $h_{\mu \nu}$ is the only ingredient with non-zero weight.  Since the perturbed Weyl scalar $\psi_4$ is linear in $h_{\mu \nu}$, it simply inherits the weight and decays as $\hat{\psi}_4 \sim v^{-1/2}$.  For the Hartle-Hawking tetrad, instead $\psi_4 \sim v^{3/2}$ since $m^\mu$ is weight $0$ but $n^\mu$ is weight $+1$.  A second example is the squared electric field strength $E^2=F_{\mu \alpha} u^\alpha F^{\mu \beta} u_\beta$ observed by an infalling observer with four-velocity $u^\mu$.  The field $A_\mu$ has weight $-1/2$, while the four-velocity $u^\mu$ has weight $+1$ \cite{Gralla:2016jfc}.  Both appear twice, so the total weight is $-1/2 - 1/2 + 1 + 1=1$.  That is, the observed squared electric field strength grows linearly in $v$ at late times on the horizon.
\subsubsection{Mode by mode weights}
The simplicity of the main result \eqref{eq:boss} belies an intricate mode-by-mode structure, which is of interest in its own right.  The mode decomposition is straightforward at the level of the Hertz potentials $\hUs$ \cite{Cohen:1974cm,Chrzanowski:1975wv,Wald:1978vm,Stewart1979Recon} we use in the computation.  The field is determined from the potentials (up to trivial changes in mass, angular momentum, and charge) by applying certain operators given in Eqs.~\eqref{eq:Pis} and \eqref{eq:morePis} below. 
Here we simply note that
\begin{subequations}\label{eq:PisIntro}
\begin{align}
\Psi & = \hat{\Upsilon}_0, \\
A_\mu & = \textrm{(weight-preserving function of } \hat{\Upsilon}_{-1}) \\ 
h_{\mu\nu} & = \textrm{(weight-preserving function of } \hat{\Upsilon}_{-2}),
\end{align}
\end{subequations}
i.e. each field is constructed from the associated potential in a manner that preserves the weight (weak or strong).
Here $\Psi$ satisfies the massless Klein-Gordon equation, $A_\mu$ satisfies the vacuum Maxwell equation, and $h_{\mu \nu}$ satisfies the vacuum linearized Einstein equation.  Additionally, $A_\mu$ and $h_{\mu \nu}$ satisfy the ingoing radiation gauge conditions $A_{\mu} \ell^\mu=0$, $h_{\mu \nu} \ell^\mu=0$ and $h_{\mu \nu}g^{\mu \nu}=0$, where $\ell^\mu$ is tangent to the ingoing principal null direction.

For near-horizon, late-time behavior, the most useful set of angular modes for our purposes is the spin-weighted spheroidal harmonics with angular frequency equal to the superradiant bound.  We denote these by $\Shz(\theta)$, with associated eigenvalue $\Khz$; see Eqs.~\eqref{eq:Sph eq} and \eqref{eq:RSKI} for details.  Each Hertz potential may be decomposed in the complete orthogonal set,
\begin{align}
\hUs(v,x,\theta,\Phi) = \sum_{\ell =|s|}^\infty \sum_{m=-\ell}^{\ell} \hU (v,x)  \, \Shz(\theta) \, e^{i m \Phi}.
\end{align}
The main result of Sec.~\ref{sec:computation} below is the near-horizon, late-time asymptotics of each mode $\hU (v,x)$, expressed as a convolution of initial data with an integration kernel $F(v,x,x')$ [Eq.~\eqref{eq:convolve}] whose asymptotics are given in closed form [Eqs.~\eqref{eq:Fsup}, \eqref{eq:Fprinc m < 0}, \eqref{eq:Fprinc m > 0}, and  \eqref{eq:Faxi}].  The axisymmetric ($m=0$) modes [Eqs.~\eqref{eq:Faxi} and \eqref{eq:paxi}] are strongly self-similar with weight $p=-\ell-2$.\footnote{Note that there is an accidental zero coefficient [Eq.~\eqref{eq:Cnls}] which gives rise to slower horizon-decay of a particular transverse derivative.  See Eq.~\eqref{eq:accident} below for the complete rates.}  The nonaxisymmetric modes split into ``principal'' and ``supplementary'' modes according to the sign of the quantity\footnote{An alternative notation $h=1/2+\sqrt{Q}$ is used primarily in the calculation.  Yet another notation $h=1/2\pm i \delta$ is common; see table I of Ref.~\cite{Gralla:2016sxp} for the relationship.}
\begin{align}
{}_s Q_{\ell m} := \frac{1}{4} + {}_s K_{\ell m} - 2m^2.
\end{align}
When $Q>0$ (supplementary modes), the modes are strongly self-similar with weight $p=-1/2-\sqrt{Q}-im$ [Eqs.~\eqref{eq:Fsup} and \eqref{eq:psup}].  On the other hand, when $Q<0$ (principal modes), the modes are only weakly self-similar, expressed as an infinite sum of terms with definite weights $p_n=-1/2 - i m  + i \sqrt{|Q|} \, \mathrm{sign} (m) (1 + 2 n)$ for $n \in \mathbb{Z}^{\geq 0}$ [Eqs.~\eqref{eq:Fprinc m < 0}, \eqref{eq:Fprinc m > 0}, and \eqref{eq:pprinc}].  (In practice, however, the first term dominates and the principal modes are strongly self-similar for all intents and purposes [Eq.~\eqref{eq:Fprinc m > 0 leading}].)  The field reconstruction \eqref{eq:PisIntro} defines modes of the tensors $\Psi, A_\mu, h_{\mu \nu}$, which inherit the weights of the associated Hertz potential.  The principal modes dominate, giving rise to the $P=-1/2$ weak weight of the total field reported above.  We may summarize the detailed mode structure by writing
\begin{align}\label{eq:show}
p = 
\begin{cases} 
-1/2 - i m  + i \sqrt{|Q|} \, \mathrm{sign} (m) (1 + 2 n) \textrm{ for } n\in \mathbb{Z}^{\geq 0}, & m \neq 0, \ Q<0 \ \textrm{(principal)} \\ 
-1/2 - \sqrt{Q} - i m, & m \neq 0, \ Q>0 \ \textrm{(supplementary)} \\
-\ell -2  & m=0 \ \textrm{ \ \ \ \ \ \ \ \ \ (axisymmetric)} \\
\end{cases} 
\end{align}
These are the critical exponents of extremal Kerr perturbations.
\section{Self-Similar Structure of Extremal Kerr Perturbations}\label{sec:computation}
Reference~\cite{Casals:2016mel} showed that the Aretakis instability is associated with a branch point at the superradiant bound frequency, which is analytically accessible using the method of matched asymptotic expansions.  In that work, only massless scalar perturbations were considered.  We now generalize to include electromagnetic and gravitational perturbations and provide additional detail revealing the self-similar structure of the fields.  We first present the perturbation formalism in tetrad-covariant form using the Geroch-Held-Penrose (GHP) approach \cite{Geroch:1973am}.  We then solve the Teukolsky equation using the Kinnersley tetrad in Boyer-Lindquist coordinates.  Last we change to a regular near-horizon tetrad and coordinates to compute the critical exponents.
\subsection{Tetrad-covariant presentation}  Consider a Newman-Penrose (NP) tetrad whose real legs $\ell^\mu$ and $n^\mu$ are (respectively) aligned with the ingoing and outgoing principal null directions.  The three types of perturbations satisfy
\begin{align}
\mathcal{O}_s [\Upsilon_s] = 0,
\end{align}
where, using the GHP notation \cite{Geroch:1973am}, the operators are written \cite{PriceThesis, Wald:1978vm}
\begin{subequations}\label{Os}
\begin{align}
\mathcal O_{0} & = \thorn'\thorn-\eth' \eth - \bar{\rho}'\thorn - \rho \thorn' + \bar{\tau} \eth + \tau \eth' \\
\mathcal O_{-1} & = \left( \thorn' - \bar \rho' \right) \left(\thorn + \rho\right)- \left(\eth'-\bar \tau\right)(\eth+\tau)  \\
\mathcal{O}_{-2} & = (\thorn'  - \bar \rho')(\thorn + 3 \rho) - (\eth' - \bar \tau)(\eth+ 3 \tau) - 3 \psi_2 .
\end{align}
\end{subequations}
Here $s=0,-1,-2$ corresponds to scalar, electromagnetic, and gravitational perturbations.  In the scalar case $s=0$, $\Upsilon_0$ is just the scalar field $\Psi$ and $\mathcal{O}_s$ is (minus one half of) the box operator $\Box=\nabla_\mu \nabla^\mu$.  In the electromagnetic ($s=-1$) and gravitational ($s=-2$) cases, $\Upsilon_s$ is a complex ``Hertz potential'' from which the field may be reconstructed by taking derivatives.  Explicitly, (using the ``ingoing'' version of the formalism) we have \cite{Cohen:1974cm,Chrzanowski:1975wv,Wald:1978vm,Stewart1979Recon}
\begin{subequations}\label{eq:Pis}
\begin{align}
\Psi & = \Upsilon_0, \\
A_\alpha & = \Pi^{(-1)}_\alpha[\Upsilon_{-1}]  + \mathrm{c.c.} \\
h_{\alpha \beta} & = \Pi_{\alpha \beta}^{(-2)}[\Upsilon_{-2}]  + \mathrm{c.c.}, 
\end{align}
\end{subequations}
where
\begin{subequations}\label{eq:morePis}
\begin{align}
     \Pi^{(-1)}_\alpha &= -\ell_\alpha (\eth + \tau) + m_\alpha (\thorn + \rho), \\
\Pi^{(-2)}_{\alpha \beta} & = \ell_\alpha \ell_\beta (\eth-\tau)(\eth+3\tau)+ m_\alpha m_\beta\left(\thorn-\rho\right)\left(\thorn+3\rho\right) \nonumber \\ & -\ell_{(\alpha}m_{\beta)} \left[ \left(\thorn-\rho+\bar\rho\right)\left(\eth+3\tau\right)+\left(\eth-\tau+\bar\tau'\right)
\left(\thorn+3\rho\right)\right].
\end{align}\label{eq:Piab}
\end{subequations}
Notice that the vector potential and metric satisfy $\ell^\mu A_\mu=0$ and $\ell^\alpha h_{\alpha \beta}=0$ as well as $g^{\alpha \beta}h_{\alpha \beta}=0$, which are known as the ingoing radiation gauge conditions.
Up to trivial changes in charge, mass, and angular momentum, all smooth electromagnetic and gravitational perturbations arise from Hertz potentials in this way.\footnote{Refs.~\cite{Wald:1973jmp,Wald:1978vm} establish this fact for gravitational perturbations.  We are unaware of a correspondingly rigorous demonstration in the electromagnetic case.  Helpful discussions of the Hertz potential formalism are found in Refs.~ \cite{PriceThesis,Ori:2002uv,Barack:2017oir}.}  By considering generic solutions for Hertz potentials we access the generic late-time behavior of perturbing fields.
\subsection{Mode Decomposition in the Kinnersley tetrad in Boyer-Lindquist coordinates}\label{sec:mode decomp}
We perform the main computation in the Kinnersley tetrad in Boyer-Lindquist coordinates,
\begin{subequations}\label{eq:Kintet}
\begin{align}
\ell^\mu &= \Big( (r^2+a^2)/\Delta, 1, 0, a/\Delta \Big), \\
n^\mu    &= \frac{1}{2 (r^2 + a^2 \cos^2\theta)} \Big( r^2+a^2, -\Delta,0,a \Big),\\
m^\mu   &= \frac{1}{\sqrt{2}(r+ia\cos\theta)} \Big( ia\sin\theta, 0,1,i/\sin\theta \Big), 
\end{align}
\end{subequations}
where $\Delta = r^2-2Mr+a^2$.
This choice is convenient because Eqs.~\eqref{Os} can be repackaged into a single ``master equation'' \cite{Teukolsky1973},
\begin{align}\label{eq:Master BL}
&\Bigg(\frac{(r^2+a^2)^2}{\Delta}-a^2 \sin^2\theta \Bigg) \frac{\pd^2 \Upsilon_s}{\partial t^2} +\frac{4 M ar }{\Delta } \frac{\pd^2 \Upsilon_s}{\pd t \pd \phi} -2 s \Big( \frac{M(r^2 -a^2)}{\Delta} -r -ia \cos\theta \Big) \frac{\pd \Upsilon_s}{\pd t} \nonumber \\
&+ \Big( \frac{a^2}{\Delta} - \frac{1}{\sin^2\theta} \Big) \frac{\pd^2 \Upsilon_s }{\pd \phi^2} - \Delta^{-s} \frac{\pd}{\pd r}\Bigg( \Delta^{s+1} \frac{\pd \Upsilon_s}{\pd r} \Bigg) - \frac{1}{\sin \theta} \frac{\pd}{\pd \theta} \Big( \sin \theta \frac{\pd \Upsilon_s}{\pd \theta} \Big) \nonumber \\ &-2s \Bigg( \frac{a(r-M)}{\Delta} + \frac{i \cos \theta}{\sin^2 \theta} \Bigg) \frac{\pd \Upsilon_s}{\pd \phi} + (s^2 \cot^2\theta -s) \Upsilon_s = 0.
\end{align}
To separate \eqref{eq:Master BL} we introduce the Laplace transform\footnote{The Laplace transform is normally defined in terms of a ``frequency'' parameter $\s$ conjugate to ``time'' $\tau$.  Here we work with $\omega=i\s$, but still use Laplace, rather than Fourier, transform.} $\tUs$ of the master function
\begin{equation}\label{eq:tilde U}
\tUs(\omega,r,\theta,\phi) = \int_{0}^{\infty} e^{i \omega t} \Us(t,r,\theta,\phi) dt 
\end{equation}
and express it as a sum over modes 
\begin{equation}\label{eq:tilde Us mode}
\tUs = \sum_{\ell m} \tUslm  e^{i m\phi} =
\sum_{\ell m} \Sh(\theta) \, \Rh(r) e^{i m \phi},
\end{equation}
where by $\sum_{\ell m}$ we mean  $\sum_{\ell = \abs{s}}^\infty\sum_{m=-\ell}^\ell$.  The angular functions 
$ \Sh(\theta)$ are spin-weighted spheroidal harmonics satisfying
\begin{align}\label{eq:Sph eq}
\Bigg[\frac{1}{\sin \theta} \frac{ d}{d \theta}\left(\sin \theta \frac{d \,}{d \theta} \right) + \left( {}_s K_{\ell m \omega } - \frac{m^2+s^2+2 m s \cos \theta}{\sin^2 \theta} - a^2 \omega^2 \sin^2 \theta -2 a \omega s \cos \theta \right) \Bigg]\Sh(\theta) = 0.
\end{align}
Imposing regularity of $\Sh$ at the poles $\theta=0,\pi$ reduces integration of the angular equation to a Sturm-Liouville eigenvalue problem. As eigenfunctions, $ \Sh(\theta) $  are complete and orthogonal in the space $L^2([ 0 , \pi], \sin\theta)$. 
The eigenvalue ${}_s K_{\ell m \omega}$ is related to the spin-weighted spheroidal eigenvalue ${}_s A_{\ell m }(a\omega)$ of \cite{Teukolsky1973} by 
\begin{equation}\label{eq:K}
{}_sK_{\ell m \omega} = {}_s A_{\ell m }(a\omega) + s(s+1) + a^2 \omega^2,
\end{equation}
which in turn is related to the quantity ${}_s E_{\ell m}(a\omega)$ of Teukolsky and Press \cite{Teukolsky:1974yv} by ${}_s E_{\ell m}(a\omega) = {}_s A_{\ell m}(a\omega) + s(s+1)$. Both ${}_s K_{\ell m \omega}$ and ${}_s E_{\ell m}(a\omega)$ are unchanged under $s \to -s$, whereas ${}_s A_{\ell m}(a\omega)$ is not, making them slightly more convenient.  
Our convention is to normalize the spin-weighted spheroidal harmonics such that 
\begin{equation}\label{eq:S normalization}
 \int_0^{2 \pi}  \! \! \int_0^\pi e^{i(m-m')\phi} \Sh(\theta) \,{}_s S^*_{\ell' m'\omega}(\theta) \sin\theta \, d\theta\, d\phi = 2\pi \delta_{\ell \ell'}\delta_{m m'}.
\end{equation}
Taking the Laplace transform of the master equation \eqref{eq:Master BL} and using the mode decomposition \eqref{eq:tilde Us mode}, it follows  that the radial functions $\Rh$ satisfy the sourced ordinary differential equation
\begin{equation}\label{eq:radeq formal}
\Eh [\Rh(r)] = \Ih(r).
\end{equation}
The operator $\Eh$ is given by
\begin{align}\label{eq:radialOP}
\Eh &= \Delta^{-s} \frac{d}{dr} \Big( \Delta^{s+1} \frac{d}{dr}\Big) + U(r), \\   \qquad & U(r):=  \left( \frac{H^2 - 2 i s (r-M)H}{\Delta} + 4 i s \omega r+2 a m \omega - {}_sK_{\omega\ell m} +s(s+1) \right)
\end{align}
with $H=(r^2+a^2)\omega - a m$, while the source $\Ih(r)$ is comprised of  mode-decomposed initial data~\cite{Campanelli:1997un}:
\begin{align}
 \Ih(r)&= \int_0^{2\pi}\! \! \int_0^\pi I_s(\omega,r,\theta,\phi) \Sh^*(\theta)\, e^{-i m \phi} \sin \theta\, d\theta \, d\phi, \label{eq:ID} \\
I_s(\omega,r,\theta,\phi) & :=  \bigg[ \left(\frac{(r^2+a^2)^2}{\Delta}-a^2 \sin^2\theta\right) \left( \pd_t \Us - i \omega \Us \right)  -2s \Big( \frac{M(r^2-a^2)}{\Delta} -r -ia\cos\theta \Big) \Us + \frac{4 M ar}{\Delta} \pd_\phi \Us\bigg]_{t=0}.\label{eq:initial data} \nonumber
\end{align}
In this approach the initial data surface is $t=0$.  We choose initial data with compact support away from the horizon, a property inherited by $\Ih$.

Using the Green function method, a particular solution of Eq.~\eqref{eq:radeq formal} is
\begin{equation}\label{eq:Rh}
\Rh(r)  = \int_{r_+}^{\infty} \tf(r,r') \Ih(r') \Delta^s(r')\, dr',
\end{equation}
where the \emph{transfer function} $\tf(\omega, r,r')$ satisfies 
\begin{equation}
\Eh [\tf(r,r') ] = \delta(r-r')/ \Delta^s(r').
\end{equation}
The freedom of homogeneous solutions can be fixed by choosing the transfer function to satisfy appropriate boundary conditions.  The physical conditions are no incoming radiation from past null infinity or the past horizon.  These can be imposed by introducing homogeneous solutions $R_{\rm{in}}$ and $R_{\rm{up}}$ defined by appropriate asymptotics \cite{Teukolsky1973,Leaver1986},
\begin{align}\label{eq:in and up BCs}
R_{\rm in} & \sim N_{\rm in} \begin{cases} \Delta^{-s} e^{- i (\omega- m \Omega_H)r_*}, &  a< M \\  \Delta^{-s - i \omega r_+} e^{2 i (\omega- m \Omega_H)/(r-r_+)}, & a=M \end{cases}, \qquad r \to r_+, \\
R_{\rm up} & \sim N_{\rm up} \frac{e^{ i \omega r_*}}{r^{2s+1}}, \qquad \qquad \qquad \qquad \qquad \qquad \qquad \qquad \ \   \, r \to \infty, \label{eq:Rup bcs}
\end{align}
where $r_* = \int (r^2+a^2)/\Delta \, dr$ and $\Omega_H = a/(2M r_+)$, with $r_+=M+\sqrt{M^2-a^2}$ the horizon radius.  We also include arbitrary normalization factors $N_{\rm in}$ and $N_{\rm up}$.  For real $\omega \notin \{0,m\Omega_H\}$, the ``in'' solution satisfies the correct boundary condition at the horizon, while the ``up'' solution satisfies the correct boundary condition at infinity~\cite{Richartz:2017qep}.   For complex $\omega$ the solutions $R_{\rm in/up}$ are defined by analytic continuation.  In this work we are particularly interested in the behavior near the non-analytic point $\omega=m\Omega_H$.  We may now build the transfer function to satisfy both conditions by taking
\begin{equation}\label{eq:trans def}
  \tf(r,r') 
  = \frac{R_{\rm in} (r_<)R_{\rm up}(r_>) }{\mathcal{ W }}, \qquad \mathcal{ W } :=  \Delta^{1+s}(r') W(r'),
\end{equation}
where $r_< \,\,(r_>)$ is the lesser (greater) of $r$ and $r'$.  Here $W= R_{\rm in} R_{\rm up}' -R_{\rm up} R_{\rm in}'$ is the Wronskian of the two homogeneous solutions.  The product $\Delta^{1+s} W$, here denoted $\mathcal{W}$, is a constant independent of $r'$.  

Taking the inverse Laplace transform (with the Bromwich integral), the time-domain master field is given by
\begin{equation}\label{eq:Ups field}
\Us(t,r,\theta,\phi)= \frac{1}{2\pi} \sum_{\ell m} e^{i m \phi} \int_{-\infty + ic}^{\infty+ic}\, e^{-i \omega t}
 \Sh(\theta) \int_{r_+}^{\infty} \tf(r,r') \Ih(r') \Delta^{s}(r') dr'
\, d\omega,
\end{equation}
where $c$ is chosen such that the integrand is holomorphic above the line $ic$.  The factor of $\Delta^s$ arises from the fact that the operator $\Eh$ is symmetric with respect to this weight.  Eq.~\eqref{eq:Ups field} is the main result of this section.
\subsection{Solution of the radial equation near $\omega=m \Omega_H$ for $a=M$}\label{sec:MAE}
In the extremal case $a=M$, the homogeneous radial equation can be solved near the superradiant bound frequency $\omega=m \Omega_H$ using the method of matched asymptotic expansions \cite{TeukolskyPress1974,bender1978advanced}.  We work in units where $M=1$ and introduce a shifted radial coordinate and frequency,
\begin{align}\label{eq:x and k}
x := r-1, \qquad k := 4 ( \omega - m \Omega_H ).
\end{align}
With these definitions, the homogeneous radial equation (\eqref{eq:radeq formal} with no source) takes the form
\begin{equation}\label{extremalradial}
\Big[x^{-2s} \frac{d}{dx} \left(x^{2s+2}  \frac{d}{dx}\right) + U\Big] \Rh = 0,
\end{equation}
where
\begin{align}\label{eq:VE}
U = &\frac{\left((x+1)^2+1\right)^2}{16 x^2}k^2+\frac{m \left(x^3+4
   x^2+8 x+4\right)+2 i s \left(x^2-2\right)}{4 x} k \\ \nonumber 
   &+\frac{1}{4} \Big(-4K+m^2 \left(x^2+4 x+8\right)+4 i m s x+4 s
   (s+1)\Big).
\end{align}
The extremal radial equation \eqref{extremalradial} is of the doubly-confluent  Heun type studied by Leaver \cite{Leaver1986}, possessing rank-1 irregular singular points at the horizon and infinity. 
To streamline the matched asymptotic expansion, we introduce \cite{Casals:2016mel}
\begin{align}\label{eq:mu and b}
\mu := \frac{ k + 2 m}{2} = 2  \omega , \qquad 
b := \frac12 + \sqrt{\frac14 + K - 2\m^2}, 
\end{align}
in terms of which the potential takes the simpler form
\begin{equation}
U = b(1-b) + s(s+1) + i \mu s x + \frac{(4+x)\mu^2 x}{4} + \frac{k^2}{4 x^2} + \frac{k(\mu-is)}{x}.
\end{equation}
For scalar fields, $b(b-1)$ is the eigenvalue of the Casimir operator of the 
$\mathrm{sl}(2,\mathbb R)$ factor in the near-horizon symmetry algebra \cite{Gralla:2015rpa,Chen:2017ofv}.\footnote{In some previous work (including our own), $b$ is denoted by $h$ and called the ``conformal weight''.  This name is in deference to holographic dualities, which we do not consider in this paper.}\footnote{Representations of $\mathrm{SL}(2,\mathbb R)$ are labeled by $b$ and the fractional part of $p$ (e.g. Ref.~\cite{Barut:1965}).  Equality of these two labels is the special case of a so-called highest-weight representation.  The $\ell m$ modes we consider are not part of highest-weight representations, and $p$ and $b$ are distinct.}  We introduce $\mu$ to enable simultaneous treatment of the axisymmetric and nonaxisymmetric cases.  In the nonaxisymmetric case one can safely set $\mu=m$ as $k\rightarrow 0$, while in the axisymmetric case one must keep $\mu=2\omega$ to satisfy the boundary conditions \eqref{eq:in and up BCs}.

We now solve the radial equation \eqref{extremalradial} under the assumption that $k \ll 1$ using the method of matched asymptotic expansions.  In this method one makes separate expansions in the near ($x \ll 1$) and far ($x \gg k$) regions before matching in the region of overlap $(k \ll x \ll 1)$.  Our main interest is in the case where one point of the transfer function is in the far-zone and one point is in the near-zone, as appropriate for studying the near-horizon effects of initial data supported away from the horizon.  In this case no explicit matching is required.
\subsubsection{Near zone}
When $x \ll 1$, Eq.~\eqref{extremalradial} becomes the ``near equation''
\begin{equation}\label{eq:near eq}
x^{-2s} \frac{d}{dx} \left(x^{2s+2} \frac{d}{dx} \Rh \right) + \Big( b(1-b) + s(s+1) +\frac{k^2}{4 x^2} + \frac{k(\mu - i s)}{x} \Big) \Rh = 0.
\end{equation}
The solution is given in terms of linearly independent Whittaker functions 
\begin{equation}\label{eq: near sols}
  \Rh^{\rm near} = a_1  \, x^{-s} W_{i \mu +s, b - 1/2}( - ik /x)+   a_2  \, x^{-s} M_{i \mu + s, b - 1/2}( - ik /x).
\end{equation}
The Whittaker functions $M_{\alpha,\beta}(z)$ and $W_{\alpha,\beta}(z)$ are multivalued functions of $z$ with branch points at $z=0$ and $z=\infty$. We choose the branch cut convention $- \pi/2 < \arg(k) < 3 \pi/2$.  The horizon boundary condition requires $a_2=0$.  Thus the ``in'' solution is given in the near-zone by
\begin{equation}\label{eq:near in}
\Rh^{\rm near,in} = x^{-s} W_{i \m +s, b-1/2}(- ik/x),
\end{equation}
where we make the simple normalization choice $a_1 =  1$.  The in solution can be determined everywhere by matching to the far-zone solutions in the region of overlap.  We will not need to perform the full matching, but we note for later reference that the large-$x$ (overlap region) behavior is
\begin{align}\label{eq:Rnear in buff asy}
\Rh^{\rm near, in}  \sim  A x^{b-1-s}  + B x^{-b-s},  \quad {\rm as} \quad x\rightarrow\infty,
\end{align}
where 
\begin{align}
\label{poodle}
A = \frac{(-i k)^{1-b} \Gamma (2 b-1)}{\Gamma(b-i \mu-s)},\qquad 
B = \frac{(-i k)^b \Gamma (1-2 b)}{\Gamma(1-b-i \mu-s)}. 
\end{align}
To isolate the leading-order $k$ dependence of the radial coefficients $A$ and $B$ when $m\neq 0$, it is convenient for subsequent calculations to introduce 
\begin{align}\label{noodle}
\hat A = \frac{\Gamma(2b)}{\Gamma(b-i\mu-s)}, \qquad 
\hat B =-\frac{\Gamma(2-2b)}{\Gamma(1-b-i\mu-s)}..
\end{align}
Under $b\to1-b$, $\Rh^{\rm near,in}$ is invariant, whereas $\hat A \to - \hat B$ and $\hat B \to -\hat A$.
\subsubsection{Far zone}
When $x \gg k$, Eq.~\eqref{extremalradial} becomes the ``far equation''
\begin{equation}\label{eq:farODE}
x^{-2s} \frac{d}{dx} \left(x^{2s+2} \frac{d}{dx} \Rh \right) +
\Bigg[ b(1-b) + s(s+1)  + i s \mu x + \frac14 (4+x) x \mu^2 \Bigg] \Rh = 0.
\end{equation}
The solutions of \eqref{eq:farODE} are given by the confluent hypergeometric functions as
\begin{equation}\label{eq:far sol}
\Rh^{\rm far}= a_3 x^{b-1-s} e^{- i \mu x/2} {}_1F_1(b+i \mu - s;2b;i \mu x) + a_4 x^{-b-s} e^{- i \mu x/2} {}_1F_1(1-b+i \mu - s;2(1-b);i \mu x).
\end{equation}
The boundary condition at infinity \eqref{eq:Rup bcs} requires
\begin{equation}\label{eq:up cond}
\frac{a_3}{a_4}= -\frac{\Gamma(2-2b)\Gamma(b-i\mu +s)}{\Gamma(1-b-i\mu+s)\Gamma(2b)} (-i\mu)^{2b-1}
=: \mathcal R.
\end{equation}
Thus, the ``up'' solution is given in the far-zone by
\begin{align}\label{eq:far up}
\Rh^{\rm far,up}= \mathcal{R} x^{b-1-s} e^{- i \mu x/2} {}_1F_1(b+i \mu - s;2b;i \mu x) + x^{-b-s} e^{- i \mu x/2} {}_1F_1(1-b+i \mu - s;2(1-b);i \mu x),
\end{align}
where we make the normalization choice $a_4=1$.  The up solution can be determined everywhere by matching to near-zone solutions in the region of overlap.  We will not need to perform the full matching, but we note for later reference that the small-$x$ (overlap region) behavior is
\begin{equation}\label{eq:Rfar up buff asy}
\Rh^{\rm far, up} \sim \mathcal{R} x^{b-1-s} + x^{-b-s}, \quad {\rm as} \quad {x \rightarrow 0}.
\end{equation} 
which we employ in the computation of the Wronskian $\mathcal{W}$. Under $b\to1-b$, $\mathcal R \to \mathcal R^{-1}$ and $\Rh^{\rm far, up} \to \mathcal R^{-1}\, \Rh^{\rm far, up}$.
\subsubsection{Transfer function}\label{subsec:TF}
To assemble the transfer function, it remains to compute the constant $\mathcal{W}$, which is related to the Wronskian of the in and up solutions.  We compute the constant easily in the buffer region via \eqref{eq:Rnear in buff asy} and \eqref{eq:Rfar up buff asy}, finding
\begin{equation}
\label{eq:wron}
\mathcal W = (-ik)^{1-b}\left( \mathcal R \hat B (-ik)^{2b-1} - \hat A\right).
\end{equation}
Under $b\to1-b$, the constant $\mathcal W$ undergoes $\mathcal W \to \mathcal R^{-1} \mathcal W$.
Using Eqs.~\eqref{eq:near in}, \eqref{eq:far up}, and \eqref{eq:wron}, we find the transfer function \eqref{eq:trans} for a source point in the far-zone and a field point in the near-zone to be 
\begin{equation}\label{eq:trans}
\tf(x,x') =\frac{ x^{-s} W_{i\mu+s,b-1/2}(-ik/x)}{ \mathcal R \hat B (-ik)^{2b-1} - \hat A} \,  (-ik)^{b-1} \, \Rh^{\rm far,up}(x')  \qquad (\textrm{for }\,\,k \ll 1, \ \ x \ll 1, \ \ x' \gg k).
\end{equation}
The transfer function $\tf$ is invariant under $b\to 1-b$. Note that the contribution  from $\Rh^{\rm far,up}$ 
is smooth in $k$ at $k=0$ when $m \neq 0$ but contains isolated poles when $m=0$.  Eq.~\eqref{eq:trans} is the main result of this subsection.
\subsection{Regular coordinates and tetrad}\label{sec:reg}
As we are ultimately interested in studying decay of perturbations near the extremal horizon where BL coordinates and Kinnersley's tetrad are ill-behaved, a new set of coordinates and tetrad are needed.  We will use the corotating coordinates introduced in Sec.~\ref{sec:scaling}, related to the Boyer-Lindquist coordinates by
\begin{align}\label{eq:coros}
v= t+r_*, \quad  \Phi = \phi + r_\sharp - \Omega_H v,
\end{align}
where
\begin{align}\label{eq:rstar and rsharp}
r_* &= 1 +x +2\left(\ln x-\frac{1}{x}\right) \qquad r_\sharp = - \frac{1}{x}.
\end{align} 
For the tetrad we rescale the $\ell$ and $n$ legs as follows,
\begin{align}\label{eq:newtetrad}
\hat{\ell}^\mu = \Lambda \ell^\mu, \qquad \hat{n}^\mu = \Lambda^{-1} n^\mu, \qquad \Lambda = v \Delta/4.
\end{align}
The $m^\mu$ and $\bar{m}^\mu$ legs are unchanged (i.e. $\hat{m}^\mu=m^\mu$).  The factor of $\Delta$ makes the tetrad regular on the future horizon, while the factor of $v$ makes it further regular in the NHEK limit (i.e., the tetrad vectors have weight $0$).\footnote{The standard Hartle-Hawking tetrad has $\Lambda= \Delta / (2(r^2+a^2))$ and is not regular in the near-horizon limit in the sense that different legs have different weights.}  This is an essential step in our analysis as it ensures that the reconstruction of fields by \eqref{eq:Pis} preserves the weight.  Ref.~\cite{Burko:2017eky} recently introduced a similar tetrad to eliminate the growth of $\psi_4$; here we see a more fundamental origin in symmetry considerations.

We now rewrite the results of the preceding section in the new coordinates and tetrad.  It follows from Eqs.~\eqref{eq:Pis} and \eqref{eq:morePis} that the Hertz potentials $\Upsilon_s$ have GHP boost-weight $s$. That is, in the new tetrad \eqref{eq:newtetrad} we have $\Upsilon_s \to \hat{\Upsilon}_s$ with
\begin{align}
\hat{\Upsilon}_s = \Lambda^{s} \Upsilon_s = \left( v \Delta/4 \right)^s \, \Upsilon_s.
\end{align}
Eq.~\eqref{eq:Ups field} now becomes
\begin{equation}\label{hUs}
\hUs(v,x,\theta,\Phi)= \frac{(v/4)^s }{4\cdot 2\pi} \sum_{\ell m} e^{i m \Phi} \int_{-\infty + ic}^{\infty+ic}\, e^{-i kv/4}
 \Sh(\theta) \int_{0}^{\infty} \htf(x,x') \Ih(x') (x')^{2s} dx'
\,  dk,
\end{equation}
where 
\begin{align}
\label{eq:gtildehat}
\htf(x,x') &:= x^{2s} e^{ i \omega r_* - i m r_\sharp} \tf(x,x')
\end{align}
Note that the hatted transfer function $\htf$ is regular on the horizon $x=0$.  
\subsection{Near-horizon, late-time asymptotics}\label{sec:LTA}
Eq.~\eqref{hUs} gives the Hertz potential in regular tetrad and coordinates with no approximation.  For late-time behavior near the horizon we can make the approximations $x \ll 1$ and $k \ll 1$ and keep only the leading non-analytic behavior.\footnote{As discussed in \cite{Casals:2016mel}, the association between small-$k$ and late times follows from the general expectation that late-time behavior is associated with the uppermost non-analytic point(s) (pole or branch point) in the transfer function.  The rule of thumb \cite{Casals:2016mel} is that one expands near the special point $k=0$, keeps the leading non-analytic term, and performs the inverse Laplace transform of that term.  Such claims can be made rigorous when the global analytic structure of the transfer function is known \cite{Doetsch1974}, but unfortunately such information is not currently available for perturbations of Kerr.  The correctness of this procedure is supported by extensive cross-checks with mathematical \cite{aretakis2012decay,Aretakis:2012bm,Lucietti2012} and numerical \cite{Lucietti:2012xr,Murata:2013daa, Burko:2017eky} work.}  For compactly supported initial data away from the horizon we may hold $x'>0$ fixed in this approximation, entailing in particular that  $x' \gg k$.  That is, the approximations of Sec.~\eqref{sec:MAE} are valid.

The expression \eqref{hUs} for the Hertz potential simplifies in important ways when $k \ll 1$.  Since factors smooth in $k$ can be replaced by their $k=0$ values, the spheroidal harmonics and their associated modes of initial data lose dependence on $\omega$ and can be commuted through the inverse Laplace transform.  To package the results we define
\begin{align}\label{eq:RSKI}
\Rfhz := {}_s R^{\rm{far,up}}_{\ell m (m/2)}, \qquad\Shz := {}_s S_{\ell m(m/2)}, \qquad \Khz := {}_s K_{\ell m(m/2)}, \qquad \Ihz := {}_s I_{\ell m(m/2)}.
\end{align}
as well as
\begin{align}\label{eq:convolve}
\hU(v,x) := \int_0^\infty (x')^{2s} \Ihz(x') \, \hF(v,x,x')dx',
\end{align}
with
\begin{equation}\label{eq:F}
    \hF(v,x,x') := 
    \frac{(v/4)^s}{4} \times \frac{1}{2 \pi}\int_{-\infty+ic}^{\infty+ic} e^{-i k v/ 4} \htf(x,x') dk.
\end{equation}
(We peel off the factor of $1/(2\pi)$ because the quantity to the right of the $\times$ sign is the inverse Laplace transform with respect to $\s=-ik$, evaluated at $\tau=v/4$.)
Now the near-horizon, late-time Hertz potential is expressed as
\begin{align}\label{eq:vtoinf}
\hUs(v \to \infty,x \rightarrow 0,\theta,\Phi) \sim \sum_{\ell m} \hU(v,x) e^{im \Phi}\Shz(\theta). 
\end{align}
This demonstrates how, at late times near the horizon, the field splits naturally into modes labeled by $\ell$ and $m$, with the time-dependence of each mode given by that of the integration kernel $\hF(v,x,x')$.  In computing $\hF$ the approximations ($k \ll 1, \ x\ll1, \ x' \gg k$) are to be used.  In particular, from Eqs.~\eqref{eq:gtildehat} and \eqref{eq:trans} we have
\begin{align} \label{eq:whose you're daddi}
\htf(x,x') = e^{i \mu/2} \frac{ \Big(x^{s +i\mu} e^{-ik/(2x)}W_{i\mu+s,b-1/2}(-ik/x) \Big) \,  \Rh^{\rm far,up}(x')  }{\mathcal R \hat B (-ik)^{2b-1} - \hat A  } (-ik)^{b-1}.
\end{align}
Depending on the value of $b$, which differs for different modes [Eq.~\eqref{eq:mu and b}], this expression may simplify further using $k \ll 1$.  To classify the modes we (following \cite{Casals:2016mel}) introduce the leading $k\rightarrow 0$ piece of $b$ by $h$,
\begin{align}
h:=\frac12 + \sqrt{\frac14 + \Khz - 2m^2}.
\end{align}
In the axisymmetric case $m=0$, the eigenvalue is just ${}_s K_{\ell 0} = {}_s K_{\ell 0 0}=\ell(\ell+1)$ \cite{TeukolskyPress1974}, giving simply that $h=\ell+1$.  For nonaxisymmetric modes, there are two natural cases, depending on whether the quantity under the square root is negative or positive.  When $h$ labels irreducible representations of $\mathrm{SL}(2,\mathbb R)$, these cases are known as ``principal'' and ``supplementary'' (respectively) \cite{Barut:1965}, and we follow Ref.~\cite{Gralla:2016sxp} in adopting that terminology here.  Numerically, one observes that supplementary modes occur for smaller values of $|m|$, with the transition to principal occurring around $|m|/\ell \approx .74$.  This transition is exact in the large-$\ell$ limit \cite{Yang:2013uba}.  To summarize, we have
\begin{align}
h \textrm{ is } \begin{cases}=1/2+i \mathpzc r, & |m| \gtrsim .74 \ell \quad \qquad \! \textrm{(principal),} \\ >1, \,\& \notin \mathbb{Z}, & 0 < |m| \lesssim .74 \ell  \quad \textrm{(supplementary),} \\ = \ell+1, & m=0 \quad  \quad \ \ \qquad \textrm{(axisymmetric),}\end{cases} \label{eq:hnotation}
\end{align}
where $\mathpzc r >0$. We now examine each case in turn.  
\subsection{Inverse Laplace Transforms}
We now perform the inverse Laplace transform for the integration kernel $\hF(v,x,x')$ \eqref{eq:F}.  We denote the usual Laplace transform using $\tau$ and $\s$ so that $U(\s) = L[u(\tau)]$ and $u(\tau)=L^{-1}[U(\s)]$ for some function $u(\tau)$ and its Laplace transform $U(\s)$.  (These map on to \eqref{eq:F} by $\s=-ik$ and $\tau=v/4$.)  We will need the following elementary Laplace transforms,
 \begin{align}
 L^{-1}[\s^N\ln (\s)] & =  \frac{ (-1)^{N+1} N!}{\tau^{N+1}},  \quad N \in \mathbb Z^{+},  \label{eq:logplace} \\
 L^{-1}[\s^c] & =  \frac{ 1 }{\Gamma(-c)\tau^{c+1}},  \quad c \notin \mathbb\ \mathbb Z^{+}. \label{eq:sc}
 \end{align}
We will also need the Laplace transform provided in Sec 5.20 of \emph{The Bateman Manuscript} \cite{bateman1954tables},
\begin{align}\label{eq:Bates Motel}
L^{-1} [ e^{\s/(2a)} \s^{- \sigma} W_{\kappa, \nu}(\s/a)] =  a^{-\kappa}  \tau^{\sigma-\kappa-1} {}_2 \tilde F_1(1/2-\kappa+\nu, 1/2- \kappa-\nu; \sigma-\kappa; -a \tau ),
\end{align}
where ${}_2\tilde F_1(a,b;c;z) = {}_2 F_1(a,b;c;z)/\Gamma(c) $ is the regularized hypergeometric function.
\subsubsection{Supplementary modes}
For supplementary modes where $h>1$, the first term in the denominator of the transfer function \eqref{eq:whose you're daddi} is subdominant as $k \rightarrow 0$.  We may therefore drop this term consistent with our approximations, giving 
\begin{align}\label{eq:supp trans}
    \htf(x,x') \sim - \hat A^{-1} e^{i m/2}  \Big(x^{s +im} e^{-ik/(2x)}W_{im+s,h-1/2}(-ik/x) \Big) \, {}_s R_{\ell m}(x') (-ik)^{h-1}. 
\end{align}
Using the inversion formula \eqref{eq:Bates Motel} with $\s=-ik$, $\kappa=im+s$, $\nu=h-1/2$, $\sigma=1-h$, $a=x$, and $\tau=v/4,$ we find that the supplementary modes carry the time dependence
\begin{equation}\label{eq:Fsup}
{}_sF_{\ell m}(v,x,x') \sim - \frac{ 
e^{im/2}{}_s R_{\ell m}(x')}{4 \hat A \, \Gamma(1-h-im-s)}\, \left( \frac{v}{4}\right)^{-h 
-im}  \Big(1+ \frac{vx}{4}\Big)^{s+ im -h}, 
\end{equation}
where we have used ${}_2 F_1(a,b;b;z)=(1-z)^{-a}$. Comparing with Eq.~\eqref{late}, we see that the supplementary modes are strongly self-similar with definite weight 
\begin{align}\label{eq:psup}
p = -h-im \qquad \rm(supplementary).
\end{align}
\subsubsection{Principal modes}
For the principal modes, the two terms in the denominator of the transfer function \eqref{eq:whose you're daddi} are of the same order as $k \rightarrow 0$.  We begin by introducing
\begin{align}
\zeta := \mathcal R\hat{B}/\hat{A},
\end{align}
in terms of which Eq.~\eqref{eq:whose you're daddi} becomes
\begin{align}\label{eq:trans prince}
\htf(x,x')  \sim -  \frac{1}{1-\zeta(-ik)^{2h-1}} \frac{ e^{i m/2}}{ \hat A}  \Big(x^{s +im} e^{-ik/(2x)}W_{im+s,h-1/2}(-ik/x) \Big) \, {}_s R_{\ell m}(x') (-ik)^{h-1}. 
\end{align}
The new feature relative to the supplementary case is the prefactor $1/(1-\zeta(-ik)^{2h-1})$.  We can express this as a geometric series in $\zeta (-ik)^{2h-1}$ when $|\zeta(-ik)^{2h-1}|<1$.  Noting $2h-1=2 i\mathpzc r$  [Eq.~\eqref{eq:hnotation}], this occurs when $|\zeta|e^{\pi \mathpzc r}<1$.  We find numerically that this condition is satisfied when $m<0$, allowing us to use the geometric series representation to perform the inverse Laplace transform term by term with Eq.~\eqref{eq:Bates Motel} (again with $\s=-ik$ and $\tau=v/4$).  In terms of $h=1/2+i\mathpzc r$, we find
\begin{align}\label{eq:Fprinc m < 0}
     {}_s F_{\ell,m<0}(v,x,x') \sim & - \frac{e^{im/2} {}_s R_{\ell m}(x') }{4 \hat A} \left(\frac{v}{4}\right)^{-1/2-i(r+m)} \nonumber\\ &\times \sum_{n=0}^\infty \zeta^{n}(v/4)^{-2inr}{}_2 \tilde F_{1}\left(h-im-s,1-h-im-s,1-h-n(2h-1)-im-s, -vx/4\right).
\end{align}
In the $m>0$ case we find numerically that $|\zeta(-ik)^{2h-1}|>1$.  We can then use a geometric series in $1/(\zeta(-ik)^{2h-1})$, giving instead that
\begin{align}\label{eq:Fprinc m > 0}
    {}_s F_{\ell,m>0}(v,x,x') \sim & \frac{e^{im/2} {}_s R_{\ell m}(x') }{4 \mathcal{R} \hat B} \left(\frac{v}{4}\right)^{-1/2+i( \mathpzc r-m)}\nonumber \\ &\times \quad \sum_{n=0}^\infty \zeta^{-n}(v/4)^{2in \mathpzc r}{}_2 \tilde F_{1}\left(h-im-s,1-h-im-s,h+n(2h-1)-im-s, -vx/4\right).
\end{align}
Recall that under $h \to 1-h$ we have $\mathcal{R} \to 1/\mathcal{R}$, $\Rfhz \to \Rfhz/\mathcal{R}$, $\zeta \rightarrow 1/\zeta$, and $\hat{A} \to -\hat{B}$.  This shows that Eqs.~\eqref{eq:Fprinc m < 0} and \eqref{eq:Fprinc m > 0} are in fact related by $h \to 1-h$,
\begin{equation}\label{eq:princely relationships}
 {}_s F_{\ell,m<0}(v,x,x') = {}_s F_{\ell,m>0}(v,x,x')\big\vert_{h\to 1-h}.
\end{equation}
Thus the principal modes are a sum over terms $n \in \mathbb{Z}^{ \geq 0}$, each with definite weight
\begin{equation}\label{eq:pprinc}
    p = -1/2 - i m + i \mathpzc r \, \mathrm{sign} (m) (1 + 2 n)   \qquad ({\rm principal}).
\end{equation}
That is, the principal modes are weakly self-similar with weight $P=-1/2$.

While the infinite series \eqref{eq:Fprinc m < 0} and \eqref{eq:Fprinc m > 0} are required for asymptotic equality as $v \rightarrow \infty$ in the strict sense, in practice the first term is numerically dominant by at least a factor of $10^3$.  If we keep only this term, the result is
\begin{align}\label{eq:Fprinc m > 0 leading}
    {}_s F_{\ell,m>0}(v,x,x') \approx \frac{e^{im/2} {}_s R_{\ell m}(x') }{4 \mathcal{R} \hat B\, \Gamma(h-im-s)} \left(\frac{v}{4}\right)^{-1/2+i(\mathpzc r-m)}\, \left(1+\frac{vx}{4}\right)^{-\frac12+s+i(m+ \mathpzc r)},
\end{align}
with the $m<0$ case given by $h \rightarrow 1-h$ (i.e. $\mathpzc r \to - \mathpzc r$).  A similar truncation was performed in the near-extremal case \cite{Gralla:2016sxp}, where higher-order terms in a parameter $\eta$ were dropped.  The time-dependence is identical under suitable identifications [see Eq.~(21) therein].
\subsubsection{Axisymmetric modes}\label{subsec:Axis}
In the axisymmetric case $m=0$, $\mu = 2\omega = k/2$ and Eq.~\eqref{eq:whose you're daddi} may be further simplified using $\mu \ll 1$.  The late-time behavior follows from the leading \textit{non-analytic} term as $\mu \rightarrow 0$ \cite{Casals:2016mel}, which is not always the leading term.  We will need the subleading expansion of the angular eigenvalue \cite{Seidel1989}
\begin{align}
{}_sK_{\ell0\omega} = \ell(\ell+1) + K_2 \mu^2 + O(\mu^4),
\end{align}
where 
\begin{equation}
    K_2 = \frac{\ell(\ell+1)(\ell(\ell+1)-1)+2\ell(\ell+1)s^2 -3 s^4}{2\ell(\ell+1)(2\ell-1)(2\ell+3)}.
\end{equation}
The ``Casimir'' $b$ is then expanded as
\begin{align}
b & = \frac12 + \sqrt{\frac14 + K - 2 \mu^2 } \\
& =  \ell +1 +b_2 \mu^2 + \orderof{\mu^4},
\end{align}
where $b_2=(K_2-2)/(1+2\ell)$. 
We will also need the expansions of Eqs.~\eqref{noodle} and \eqref{eq:up cond}:
\begin{subequations}\label{eq:Laurents}
\begin{align}
\hat A &= \frac{(2\ell+1)!}{(\ell-s)!} + \orderof{\mu},\\
\hat B &= \frac{(-1)^{\ell+1+s} (\ell+s)!}{ 2 b_2(2 \ell)!  (- i\mu)}  + \orderof{1},\\
\mathcal R &= \frac{(-1)^{\ell+s+1} (\ell+s)!(\ell-s)!}{2b_2 (2\ell)! (2\ell+1)!} (- i \mu)^{2b-2}\left(1+\orderof{\mu}\right).
\end{align}
\end{subequations}

 To derive the late-time decay of the axisymmetric modes, we can no longer rely on Eq.~\eqref{eq:Bates Motel} for the inverse Laplace transform due to the presence of $i\mu$ in the first index of the Whittaker function in \eqref{eq:whose you're daddi}. Instead, we shall make use of the asymptotics of the axisymmetric near in function~\cite{nist}
 \begin{equation}\label{eq:RinNEAR asy}
{}_s R^{\rm near, in}_{\ell \omega} \sim  x^{-s} e^{i\mu/x} (-2 i\mu/x)^{i\mu+s} \sum_{n=0}^{\infty}  E_{n\ell s}(\mu) \cdot (2i\mu/x)^{-n}, \quad x \to 0,
\end{equation}
where
\begin{equation}
 E_{n \ell s}(\mu) = (b- i \mu-s)_n (1-b-i\mu -s)_n/ n!.
\end{equation}
Here $(a)_n=\Gamma(a+n)/\Gamma(a)$ is the Pochhammer symbol.  
Equation~\eqref{eq:RinNEAR asy} is valid as an asymptotic series for $x/\mu \ll 1$.  Our results for the inverse Laplace transform will similarly be valid only for $x v \ll 1$.  This is in contrast to the nonaxisymmetric cases, where we performed the full inverse Laplace transform, valid for any $x \ll 1$ and $v \gg 1$.  In the axisymmetric case, we demonstrate the self-similarity in the form of an asymptotic series in $xv \ll 1$.  In the remainder of this section, the symbol $\sim$ denotes asymptotic equality as $x \to 0$, $\mu \to 0$, $x/\mu \to 0$ (and fixed $x'$).

Noting that the $\mu \rightarrow 0$ asymptotics of $\Rfhz$ [Eq.~\eqref{eq:far up}] are equivalent to its $x \rightarrow 0$ asymptotics [Eq.~\eqref{eq:Rfar up buff asy}], and using Eqs.~\eqref{eq:Laurents} and \eqref{eq:RinNEAR asy}, the transfer function \eqref{eq:whose you're daddi} becomes
\begin{align}\label{eq:tfh axi}
{}_s \hat{\tilde{g}}_{\ell\omega}(x,x') &\sim (- 2 i \mu)^{ i \mu + s+\ell} \mathcal F_{\ell s}(x') \sum_{n=0}^\infty  E_{n\ell s}( \mu) \left( \frac{x}{2 i \mu }\right)^n,
\end{align}
 where
\begin{equation}\label{eq:curly F}
\mathcal F_{\ell s}(x') := \begin{cases}\displaystyle \frac{b_2 (x'-2b_2)}{x'(2b_2^2 -1) }\,\,, & \ell =0 , \\
\displaystyle
  -\frac{(\ell-s)!}{(1+2 \ell)!} (x')^{-1-\ell-s}, & \ell >0.
  \end{cases}
\end{equation}
We may further expand in $\mu$ as
\begin{align}
(-2 i\mu)^{i\mu} & = 1 + i \mu \ln(-2 i \mu) + O((\mu \ln \mu)^2),
\end{align}
and
\begin{align}\label{eq:EjHH}
 E_{j\ell s}( \mu) = 
 \begin{cases}E^\leq_{j \ell s} + O(\mu) \,\,, & j \leq \ell + s, \\
  E^>_{j \ell s} \,\, \mu + O(\mu^2), & j > \ell +s,
 \end{cases}
\end{align}
where
\begin{align} 
E^\leq_{j \ell s} &=\frac{(-1)^j (\ell+s)! (\ell-s+j)!}{ j!\, (\ell-s)! (\ell+s-j)!}, \\
E^>_{j\ell s} &= \frac{i (-1)^{1+\ell+s} (\ell+j-s)!(\ell+s)!(j-\ell-s-1)!}{j!\, ( \ell-s)!}.
\end{align}
Plugging back in to Eq.~\eqref{eq:tfh axi} and keeping only leading non-analytic parts of each term $n$ in the asymptotic series, we find
{\small
\begin{align}\label{eq:axi late-time trans}
{}_s \hat{\tilde{g}}_{\ell\omega}(x,x') \sim &\mathcal F_{\ell s}(x')
(-2i\mu)^{s+\ell+1}\Bigg(
   -\frac12 \ln(-2i\mu) \sum_{n=0}^{s+\ell} E^{\leq}_{n\ell s}\left(\frac{x}{2 i\mu}\right)^{n}+ (4i)^{-1} \, E^>_{(s+\ell+1)\ell s}\left(\frac{x}{2 i\mu}\right)^{s+\ell+1} \, (-2i\mu)\ln(-2i\mu)\nonumber \\
   & \qquad \qquad \qquad \qquad \qquad \qquad + \sum_{n=\ell+s+2}^\infty  E^>_{n\ell s}\left(\frac{x}{2 i\mu}\right)^{n}
  \Bigg) + \textrm{(terms smooth in $\mu$)}.
\end{align}}
The terms smooth in $\mu$ do not contribute to the Laplace transform at late times.  Using Eq.~\eqref{eq:logplace} and \eqref{eq:sc} (together with $\s=-ik$ and $\tau=v/4$) we compute the inverse Laplace transform \eqref{eq:F} term by term in the asymptotic series, the result being  
\begin{align}
     {}_s F_{\ell 0}(v,x,x') \sim &\frac{1}{4}\mathcal{F}_{\ell m} (v/4)^{-\ell-2}
     \bigg( (-2)^{-1}\sum_{n=0}^{\ell+s} E^\leq_{n \ell s} (-1)^{s+\ell}(1+\ell+s-n)!(vx/4)^n \nonumber \\
     & \qquad \qquad \qquad \qquad \qquad + \frac{1}{v} \frac{(-1)^{s+\ell+1}4^{\ell-1}}{4i} E^>_{(s+\ell+1)\ell s} (xv/4)^{s+\ell+1} \nonumber \\
     & \qquad \qquad \qquad \qquad \qquad + \sum_{n=\ell+s+2}^\infty \frac{E^>_{n \ell s} (-vx/4)^n}{(n-s-\ell-2)!} \bigg). \label{eq:wowza}
\end{align}
  Notice that term on the middle line (originating from the last term on the first line of \eqref{eq:axi late-time trans}) is subleading as $v \rightarrow \infty$.  Dropping this term gives the leading self-similar form,
 \begin{align}\label{eq:Faxi}
    {}_s F_{\ell 0} (v,x,x') \sim &\frac{1}{4}\mathcal{F}_{\ell m}(x')\, (v/4)^{-\ell-2}\,\sum_{n=0}^{\infty} C_{n \ell s} (vx/4)^n, 
 \end{align}
 where 
  \begin{align}\label{eq:Cnls}
     C_{n \ell s} :=
     \begin{cases}
\displaystyle
     (-1)^{s+\ell+n-1}\frac{ (1+s+\ell-n)!(\ell+s)! (\ell-s+n)!}{ n!\, (\ell-s)! (\ell+s-n)!} \quad &n \leq \ell +s,\\
     0 \quad &n =\ell +s+1,\\
\displaystyle
    i (-1)^{1+\ell+s+n}\frac{(\ell+n-s)!(\ell+s)!(n-\ell-s-1)!}{n!\, ( \ell-s)!}\quad &n>\ell+s+1.
     \end{cases}
 \end{align}
From Eq.~\eqref{eq:Faxi} we see that (at least asymptotically for small $vx$) the axisymmetric modes are strongly self-similar with definite weight
\begin{equation}\label{eq:paxi}
    p = -\ell -2 \qquad (\mathrm{axisymmetric}).
\end{equation}

The anomalous $0$ in Eq.~\eqref{eq:Cnls} indicates that the leading self-similar behavior \eqref{eq:Faxi} has vanishing $x$-derivative of order $1+\ell+s$ on the horizon.  This accidental zero gives rise to slower decay for that particular derivative.  This decay arises from the subleading term in the middle line of \eqref{eq:wowza}, which is $1/v^{-2+s}$.  To wit, the decay of each derivative is
\begin{equation}\label{eq:accident}
\pd_x \left[{}_s F_{\ell 0}(v,x,x')\right]_{\mathcal H} \sim  Z_{n \ell s}(x')   \begin{cases} v^{-2+s} \,\,, & n  =\ell +s +1 , \\
  v^{-2-\ell+n}, & \mathrm{otherwise}.
  \end{cases}
\end{equation}
The function $ Z_{n \ell s}(x') $ may be straightforwardly computed using expressions \eqref{eq:EjHH} and \eqref{eq:curly F}.  These rates are inherited by each mode of the field via Eq.~\eqref{eq:convolve}.  This generalizes the decay/growth results presented in the spin-$0$ case in Ref.~\cite{Casals:2016mel}.
\section{Outlook}\label{sec:outlook}
We now place this work in context and discuss some future directions.  One interesting direction would be toward more mathematical rigor.  Despite its enormous track record of success, the method of matched asymptotic expansions does not seem to lend itself to rigorous justification, and in particular we have given no rigorous proof of the asymptotic estimates it entails.  Furthermore, we have not investigated the global structure of the transfer function in the complex plane, and thereby have to \textit{assume}, rather than demonstrate, that the branch point we study indeed gives the late-time behavior.  While numerical, analytical, and mathematical cross-checks leave no doubt about the validity of the results, a more rigorous derivation may reveal additional interesting features and more sharply delineate the scope of the results. 

Second, it would be desirable to extend this analysis to horizon-penetrating initial data.  Such data changes the decay rate of axisymmetric perturbations \cite{Aretakis:2012ei,Lucietti:2012sf,Aretakis:2012bm}, while the situation for nonaxisymmetric perturbations is not yet clear \cite{Burko:2017eky,Hadar:2017ven}. We are confident on physical grounds that self-similarity will emerge in the horizon-penetrating case as well, but new techniques may be required to demonstrate it.  We note, however, that once self-similarity is demonstrated and the associated critical exponents are determined, the decay/growth rate for any tensorial quantity follows immediately by the methods of Sec.~\ref{sec:scaling}.  In particular, we expect all scalars to decay.

It would also be interesting to study the approach to critical behavior by generalizing to \textit{nearly} extremal black holes.  The basic properties can be understood from our previous work with A. Zimmerman \cite{Gralla:2016sxp}.  In that work, we demonstrated that Aretakis behavior occurs transiently over a time of order $1/\kappa$, where $\kappa$ is the surface gravity of the black hole.  The fields take the characteristic self-similar form seen here, arising in the near-extremal context as a coherent superposition of near-horizon quasi-normal modes.   
One is thereby tempted to view the self-similarity as a collective self-organization occurring over a timescale of $1/\kappa$.  This ``correlation time'' diverges in the critical limit, as expected for a critical phenomenon.  

Another very interesting direction would be to consider non-linear perturbations.  With generic initial data, a precisely extremal black hole will immediately become non-extremal when the first radiation crosses the horizon, a non-linear effect not captured by our analysis.  However, working in the near-extremal case, one can choose the amplitude of the perturbation small enough to preserve near-extremality.  There is unlikely to be a large breakdown of linear theory since scalar invariants decay.  However, the decay is a slow polynomial, giving more time for non-linear mode couplings to develop, perhaps leading to turbulence \cite{Yang:2014tla}.  We hope that the framework presented here, wherein a symmetry principle organizes the calculation, will serve as a foundation for exploring the exciting prospect of non-linear turbulent critical phenomena near rapidly rotating black holes.

Critical phenomena are typically associated with universality: the same exponents apply to many different systems.  Hints of such universality are already seen here.  First, we found the same weak exponent $P=-1/2$ for all three types of perturbations considered (scalar, electromagnetic, gravitational).  Second, charged complex scalar perturbations of charged black holes behave in a precisely analogous manner, sharing the same weak exponent $P=-1/2$ together with much, if not all, of the detailed structure \eqref{eq:show} \cite{Zimmerman:2016qtn}.\footnote{The charge $q$ of the field plays the role of the azimuthal number $m$.}  More generally, it is known that extremal horizons generically enjoy near-horizon symmetry enhancements \cite{Kunduri:2007vf}.  It seems that self-similarity of perturbing fields is a generic property of physics near extremal horizons.  Perhaps the critical exponents can be organized into universality classes, of which $P=-1/2$ is merely the first to be discovered.

\section*{Acknowledgements}
We thank Stefanos Aretakis, Lior Burko, Marc Casals, Gaurav Khanna, Alex Lupsasca and Aaron Zimmerman for helpful conversations. This work was supported by NSF grant 1506027 to the University
of Arizona.
\appendix 
\section{Self-similar tensor fields in NHEK}\label{app:def weight tensors}\label{appendix}
A  rank $(r,s)$ tensor field $\bar W^{\alpha_1 \cdots \alpha_r}_{\beta_1 \cdots \beta_s}$ on the NHEK spacetime is said to be self-similar with weight $p$ if 
\begin{equation}\label{eq:def weight eq}
    \Lie_{H_0}\bar W^{\alpha_1 \cdots \alpha_r}_{\beta_1 \cdots \beta_s}= p \,\bar W^{\alpha_1 \cdots \alpha_r}_{\beta_1 \cdots \beta_s},
\end{equation}
where $H_0^\mu \pd_\mu = \bar v \pd_{\bar v} - \bar x\pd_{\bar x}$ is the generator corresponding to infinitesimal dilations.  (In this section we will not raise and lower indices, making the notation $\bar W^{\alpha_1 \cdots \alpha_r}_{\beta_1 \cdots \beta_s}$ unambiguous.)  Writing out the Lie derivative using the partial derivative operator gives \cite{Wald1984}
\begin{equation}\label{eq:Lie def}
     \Lie_{H_0} \bar W^{\alpha_1 \cdots \alpha_r}_{\beta_1 \cdots \beta_s}
     = H_0^\mu \pd_\mu \bar  W^{\alpha_1 \cdots \alpha_r}_{\beta_1 \cdots \beta_s}
     -\sum_{i=1}^r \bar W^{\alpha_1 \cdots \gamma \cdots \alpha_r}_{\beta_1 \cdots \beta_s} \pd_\gamma H_0^{\alpha_i}+\sum_{j=1}^s \bar  W^{\alpha_1 \cdots \alpha_r}_{\beta_1 \cdots \gamma \cdots \beta_s} \pd_{\beta_j} H_0^\gamma.
\end{equation}
To find solutions we introduce coordinates $\bar{X}^\alpha=(\bar V, \bar X, \bar \theta, \bar \Phi)$ where
\begin{align}
\bar{V} = \bar v, \qquad \bar{X}= \bar x \bar v
\end{align}
such that 
\begin{equation}\label{eq:H0X}
    H_0^\mu \pd_\mu =  \bar{V}\pd_{\bar{V}}.
\end{equation}
We begin with a $(1,0)$ tensor field $\bar W^\alpha$ before tackling the general case.  
In this case \eqref{eq:Lie def} reads
\begin{align}\label{eq:LieH0 vec}
    \Lie_{H_0}\bar W^\alpha & = \bar{V}\pd_{\bar{V}} \bar W^\alpha -\bar W^\beta \pd_\beta \left(\bar{V}\pd_{\bar{V}}^\alpha \right) \\
    & = \bar{V}\pd_{\bar{V}} \bar W^\alpha - \bar{W}^{\bar{V}} \pd_{\bar{V}}^\alpha 
\end{align}
Thus the  $\bar{V},\bar{X}$ coordinate components of $\bar{W}^\alpha$ satisfy
\begin{align}
\bar{V} \pd_{\bar{V}} \bar{W}^\alpha = \begin{cases} 
p \bar{W}^\alpha, & \alpha \neq \bar{V} \\
(p+1) \bar{W}^\alpha, & \alpha = \bar{V}
\end{cases} ,
\end{align}
with solution
\begin{align}
\bar{W}^\alpha = f^\alpha(\bar X,\bar{\theta},\bar{\Phi}) \begin{cases} 
\bar{V}^p , & \alpha \neq \bar{V} \\
\bar{V}^{p+1} & \alpha = \bar{V}
\end{cases}.
\end{align}
We can repackage this as
\begin{equation}
    \bar W^\alpha = \bar V^{p+\Gamma[\alpha]} f^{\alpha} (\bar{X},\bar \theta,\bar \Phi)
\end{equation}
where the exponent $\Gamma[\alpha] =1 $ for $\alpha = \bar V$ and is zero otherwise.  Using the relation $\Lie_{H_0}\left(\bar W_\alpha  \bar W^\alpha \right) = 2p \bar W_\alpha \bar W^\alpha$, the general solution for a $(0,1)$ tensor $ \bar W_\alpha$ is found to be $ \bar W_\alpha = \bar{V}^{p - \Gamma[\alpha]} f_\alpha(\bar X,\bar\theta,\bar\Phi)$. More generally, the solution of \eqref{eq:def weight eq} for an $(r,s)$ tensor is given by
\begin{equation}
   \bar  W^{\alpha_1 \cdots \alpha_r}_{\beta_1 \cdots \beta_s} = \bar{V}^{p + \Gamma[\alpha_1, \cdots, \alpha_r]-\Gamma[\beta_1, \cdots, \beta_s]} f^{\alpha_1 \cdots \alpha_r}_{\beta_1 \cdots \beta_s}(\bar X,\bar\theta,\bar\Phi),
\end{equation}
where now $\Gamma[\gamma_1, \cdots, \gamma_i]$ returns the number of $\bar{V}$ indices (e.g. $\Gamma[\bar{V} \bar{X} \bar{V}]$=2).  

To go back to the NHEK ingoing coordinates $\bar{v},\bar{x}$ coordinates, we use the transformation matrix
\begin{equation}
    \frac{\pd \bar{x}^\mu}{\pd \bar{X}^\a} = 
    \begin{bmatrix}
    1 & 0 & 0 & 0 \\
    -\bar{x}/\bar{v} & 1/\bar{v} & 0 &0 \\
    0&0&1&0 \\ 0&0&0&1
    \end{bmatrix}
\end{equation}
giving the general form 
\begin{equation}\label{eq:Wapp1}
     \bar  W^{\alpha_1 \cdots \alpha_r}_{\beta_1 \cdots \beta_s} =  \bar{v}^{p+\bar N[\alpha_1, \cdots, \alpha_r; \beta_1, \cdots, \beta_s]}  g^{\alpha_1 \cdots \alpha_r}_{\beta_1 \cdots \beta_s}(\bar v\bar x,\bar \theta,\bar\Phi),
\end{equation}
where $g$ is smooth in its arguments and $\bar N=({\rm \# \,of\, upper}\, \bar v\, {\rm indices}-{\rm \#\, of\, upper}\, \bar x\, {\rm indices})+ ({\rm \# \,of\, lower}\, \bar x\, {\rm indices}-{\rm \#\, of\, lower}\, \bar v\, {\rm indices})$.  If the self-similar field $\bar{W}$ descends from a Kerr field $W$ via Eq.~\eqref{W}, we may compute the near-horizon, late-time behavior of the parent field $W$ by changing to unbarred corotating coordinates.  Using Eq.~\eqref{eq:corotating}, we find 
\begin{equation}\label{eq:Wapp2}
      W^{\alpha_1 \cdots \alpha_r}_{\beta_1 \cdots \beta_s} =  v^{p+N[\alpha_1, \cdots, \alpha_r; \beta_1, \cdots, \beta_s]}  h^{\alpha_1 \cdots \alpha_r}_{\beta_1 \cdots \beta_s}(vx,\theta,\Phi) \textrm{ \ \ \ \ as $v \to \infty, \ x \to 0$},
\end{equation}
where $h$ is smooth in its arguments and $N=({\rm \# \,of\, upper}\, v\, {\rm indices}-{\rm \#\, of\, upper}\,x\, {\rm indices})+ ({\rm \# \,of\, lower}\, x\, {\rm indices}-{\rm \#\, of\, lower}\,v\, {\rm indices})$.
\bibliography{MyReferences.bib}
\end{document}